\documentclass[preprint,final,aps,prb,amsfonts,letterpaper,citeautoscript]{revtex4-1}

\usepackage{CJK} 
\usepackage{graphicx, multirow,bm, amsmath}
\usepackage{hyperref}
\usepackage{bbm}
\usepackage[usenames,dvipsnames]{color}
\usepackage[version=4]{mhchem}

\usepackage[final]{changes}
\definechangesauthor[name={Fei}, color=blue]{Fei}
\definechangesauthor[name={Wes}, color=RoyalPurple]{Wes}
\definechangesauthor[name={VO}, color=teal]{VO}
\definechangesauthor[name={Yi}, color=green]{Yi}

\setremarkmarkup{[#1 : #2]}

\begin{document}
	\begin{CJK*}{UTF8}{bsmi} 
	\title{Compressive sensing lattice dynamics. II. Efficient phonon calculations and long-range interactions}
    \author{Fei Zhou(周非)} \email{fzhou@llnl.gov}
	\author{Babak Sadigh}
	\author{Daniel \AA{}berg}
	\affiliation{Physical and Life Sciences Directorate, Lawrence Livermore National Laboratory, Livermore, California 94550, USA}
	\author{Yi Xia}
	\affiliation{Department of Materials Science and Engineering, Northwestern University, Evanston, IL 60201, USA}
	\author{Vidvuds Ozoli\c{n}\v{s}}
	\affiliation{Department of Applied Physics, Yale University, New Haven, CT 06520, USA} 
	\affiliation{Yale Energy Sciences Institute, Yale University, West Haven, CT 06516, USA} 

		\date{\today} 

\begin{abstract}
We apply the compressive sensing lattice dynamics (CSLD) method to calculate phonon dispersion for crystalline solids. While existing methods such as frozen phonon, small displacement, and linear response are routinely applied for phonon calculations, they are considerable more expensive or cumbersome to apply to certain solids, including structures with large unit cells or low symmetry, systems that require more expensive electronic structure treatment, and polar semiconductors/insulators. In the latter case, we propose an approach based on a corrected long-range force constant model with proper treatment of the acoustic sum rule and the symmetric on-site force constant matrix.
Our approach is demonstrated to be accurate and efficient for these systems through case studies of NaCl, CeO$_2$, Y$_3$Al$_5$O$_{12}$ and La$_2$Fe$_{14}$B.
\end{abstract}
\maketitle
\end{CJK*}

\section{Introduction}
With the advent of efficient density-functional theory (DFT) based methods for solving the electronic ground state under the Born-Oppenheimer approximation, several {\it ab initio\/} methods for calculating the harmonic force constants or force constant matrix (FCM) of crystalline solids have been proposed, such as the frozen phonon approach \cite{Wendel1978PRL950,Ho1984PRB1575}, the direct or supercell small displacement method \cite{Kunc1982PRL406,Parlinski1997PRL4063}, the density-functional perturbation theory (DFPT) \cite{Baroni2001RMP515}, and the compressive sensing lattice dynamics (CSLD) method\cite{Zhou2014PRL185501,Soderlind2015SR15958}.  Due to these developments, {\it ab initio\/} determination of the harmonic phonon dispersion curves and phonon mode Gr\"uneisen parameters has become routine and readily available in many software packages (for a recent review see Ref.~\onlinecite{Wang2016CM16006}).

Even though there is in principle no obstacle, phonon calculations still require a significant amount of effort in practice, especially for solids with large unit cells or low symmetry. Consider the worst case senario: a system with a large primitive cell ($N_a$ atoms) and no symmetry. Under the ``shortsightedness'' assumption, each atom interact appreciably with up to $N_n$ atoms. The total number of non-zero FCM elements is $\sim \frac{9}{2} N_a N_n$. In a supercell of $N_{\mathrm{p}}$ primitive cells, each DFT calculation returns information of $3 N_{\mathrm{p}} N_a -3$ force components. Assuming a ratio $g>1$ for the number of training data points to the number of unknowns, one needs at least
\begin{eqnarray}
\label{eq:n-cell}
\frac{9}{2}g N_a N_n / \left( 3 N_{\mathrm{p}} N_a -3 \right) \approx \frac{3}{2} g N_n  / N_{\mathrm{p}}
\end{eqnarray}
supercell calculations. In the limit of very large unit cells ($N_a \rightarrow \infty$), $N_\text{p} \rightarrow 1$, the above number $3N_n$ (assuming $g=2$) remains finite as long as the atomic interactions are short-ranged. By taking advantage of the sparsity or ``short-sightedness'' of the force constant, CSLD allows one to stay close to the above lower limit. In contrast, the direct method, arguably the most widely used approach, requires $\sim 3 g N_a $ supercell calculations to displace the three coordinates of each atom in the plus and minus directions ($g=2$) in the worst case, a number that grows linearly with the size of the unit cell.

Part I of this series contains technical details of CSLD while Part II focuses on phonon calculations. First, we will clarify the theory of long-range Coulomb interactions in gapped systems.

\begin{figure}[b]
	\includegraphics[width = 0.96 \linewidth]{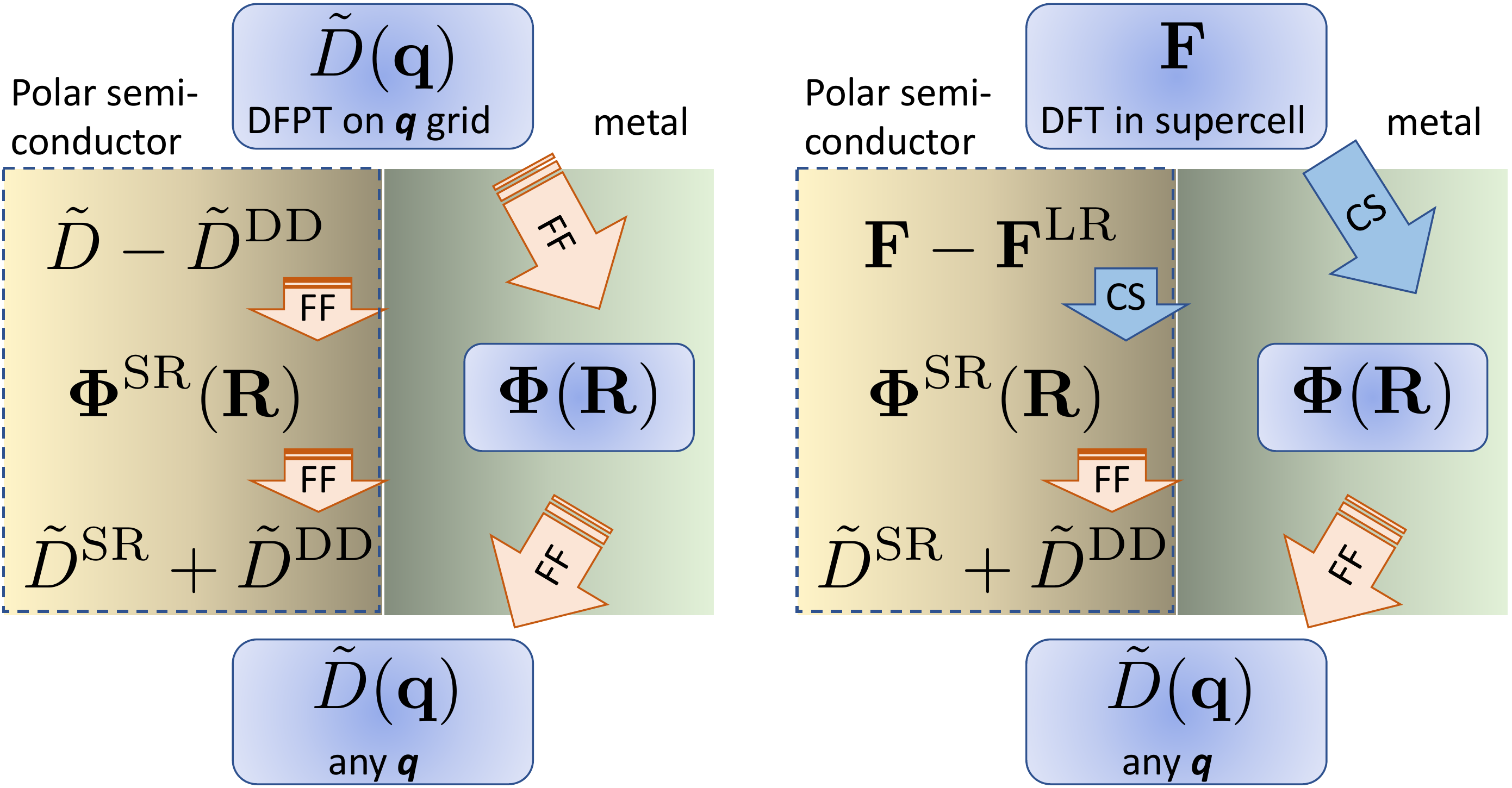}
	\caption{Comparison of phonon calculations using DFPT (left) and CSLD (right). FF=Fourier transformation.}
	\label{fig:phonon-steps}
\end{figure}

\section{Long range interactions}

In polar solids with a band gap, the long-range Coulomb interactions have to be treated with care, as they give rise to the well-known physical effect of longitudinal/transverse-optical (LO-TO) splitting \cite{Baroni2001RMP515}. In practical calculations, the force constants have to be cut off at some maximum interaction distance, but the LO-TO splitting is immediately lost at any finite cut-off. The problem is particularly problematic for CSLD, since the effectively infinite number of interacting atoms ($N_n \rightarrow \infty$) is incompatible with the very assumption of compressive sensing \cite{Nelson2013PRB035125,Nelson2013PRB155105}: sparsity in the unknown parameters (force constants).

In the direct method, the LO-TO splitting problem is treated with the help of the non-analytical part of the long-range dynamical matrix $\tilde D^{\text{NA}}(\mathbf{q} \rightarrow 0)$ in the long wavelength limit \cite{Cochran1962JPCS447}. At finite $\mathbf{q}$, $\tilde D^{\text{NA}}$ is mixed in with an interpolation scheme,  using either the semi-empirical Guassian smoothing function by Parlinski and co-workers \cite{Parlinski1998PRL3298}, or the mixed-space approach by Wang and co-workers \cite{Wang2010JPCM202201}. Detailed discussions can be found in Ref.~\onlinecite{Wang2016CM16006}.

\subsection{Range-separation in real space} \label{sec:separation}

Our approach is based on the idea of separating the long-range interactions and the residual short-range ones in real space (Fig.~\ref{fig:phonon-steps}). 
We divide the force constant matrix between atoms $a$ (site $\kappa$ in unit cell 0) and $b$ ($\kappa'$ in cell $\mathbf{R}$) into the long-range dipole-dipole term and a residual, presumptively short-range one:
\begin{align} \label{eq:SR}
\Phi_{ab} =  C_{a b}^{\mathrm{DD}} + \Phi_{ab}^{\text{SR}}.
\end{align}
The atomic forces are also range-separated:
\begin{align} \label{eq:force-separation}
\vec{F}_a = \vec{F}_a^\text{SR} + \vec{F}_a^\text{LR} = \vec{F}_a^\text{SR} -\sum_b \tilde{C}_{a b}^{\prime \mathrm{DD}} \cdot \vec{u}_b,
\end{align}
where the summation is over all atoms in a given supercell, and $\tilde{C}^{\prime \mathrm{DD}}$ is the   Fourier-transformation of ${C}^{\mathrm{ DD}}$
at $\mathbf{q}=0$ (see below). The prime indicates that $\tilde{C}^{DD}(\mathbf{q}=0)$ is evaluated for the considered supercell, not the primitive cell. Next, the short-range force constants $\{\Phi_{ab}^{\text{SR}}\}$ are fitted to $\{\vec{F}_a^\text{SR}\}$ with CSLD. Finally, the total dynamical matrix   
\begin{align}
\tilde D_{\kappa \kappa'}=\tilde C_{\kappa \kappa'}/\sqrt{M_\kappa M_{\kappa'}}  = \tilde D^{\text{SR}}_{\kappa \kappa'} + \tilde D^{\text{DD}}_{\kappa \kappa'}
\end{align}
is simply the sum of the short/long-range contributions at any wavelength, without having to resort to a mixing scheme. In the following, we will refer to both $\tilde C$ and $\tilde D$ as the dynamical matrix for brevity.
The CSLD procedure for separating force constants in real space is outlined in Fig.~\ref{fig:phonon-steps} and compared schematically to reciprocal space range-separation in DFPT \cite{Gonze1997PRB10355}.

\subsection{Long-range force constant ansatz}

Motivated by the well-known non-analytic dynamical matrix $\tilde{C}^{\mathrm{NA}}$ responsible for LO-TO splitting  \cite{Baroni2001RMP515}, 
\begin{align}
\tilde{C}_{\kappa i, \kappa' j}^{\mathrm{NA}}(\mathbf{q} \rightarrow 0) = \frac{4\pi}{\Omega_0} \frac{(q_k Z^*_{\kappa,ki}) (q_l Z^*_{\kappa',lj})}{\mathbf{q}\cdot \epsilon \cdot \mathbf{q}},  
\label{eq:C-NA}
\end{align}
the following ansatz for the dipole-dipole force constant matrix for atoms $0, \kappa$ and $\mathbf{R}, \kappa'$ was introduced in order to reproduce eq.~(\ref{eq:C-NA}) in the long-wavelength limit\cite{Gonze1994PRB13035, Gonze1997PRB10355}:
\begin{eqnarray}
C_{\kappa i, \kappa' j}^{\mathrm{DD}}(0,\mathbf{R}) &=&  \sum_{i^{\prime}j^{\prime}}  Z^{*}_{\kappa ,ii^{\prime}} Z^{*}_{\kappa', jj^{\prime}} V^{\mathrm{DD}}_{\kappa i', \kappa'j'}(0,\mathbf{R}),
\label{eq:CDD-prev} \\
V^{\mathrm{DD}}_{\kappa i, \kappa'j}(0,\mathbf{R}) &= &  (\mathrm{det} \epsilon  )^{-1/2}  \left[ \frac{(\epsilon^{-1})_{ij}}{{D^3}}-3 \frac{\Delta_{i }\Delta_{j}} {{D^5}}  \right], \nonumber
\end{eqnarray}
where $Z^{*}_{\kappa,ij}$ and  $\epsilon_{ij}$ are the Born effective-charge and ion-clamped dielectric permittivity tensors, respectively, $\Delta_{i}=\sum_{j} (\epsilon^{-1})_{ij}d_{j}$ is the conjugate of the vector $\mathbf{d}=\mathbf{r}_{\kappa'}+   \mathbf{R} -\mathbf{r}_{\kappa}$, and $D=\sqrt{\mathbf{\Delta}\cdot\mathbf{d}}$.

The set of FCMs in Eq.~(\ref{eq:CDD-prev}) should obey the constraints  discussed in Part I. The  acoustic sum rule (ASR) due to translational invariance gives the on-site FCM \cite{Pick1970PRB910, Gonze1997PRB10355}
\begin{eqnarray}
C_{\kappa i, \kappa j}^{\mathrm{DD}}(0,0) &=& -\sum_{\kappa', \mathbf{R} \neq \kappa, 0} C_{\kappa i, \kappa' j}^{\mathrm{DD}} (0, \mathbf{R}).
\label{eq:CDD-onsite}
\end{eqnarray}
Additionally, $C_{\kappa i, \kappa j}^{\mathrm{DD}}(0,0)$ is symmetric (in indices $i, j$ unless otherwise noted), and therefore
\begin{align}
\sum_{\kappa', \mathbf{R} \neq \kappa, 0} \left[ C_{\kappa i, \kappa' j}^{\mathrm{DD}} (0, \mathbf{R}) -   C_{\kappa j, \kappa' i}^{\mathrm{DD}} (0, \mathbf{R})\right]=0.
\label{eq:symmetric}
\end{align} 
Turning to the reciprocal space, the long-range dynamical matrix is \cite{Pick1970PRB910, Gonze1997PRB10355}
\begin{eqnarray}
\tilde{C}^{\mathrm{DD}}_{\kappa i, \kappa' j} (\mathbf{q})
&=& \hat{C}^{\mathrm{DD}}_{\kappa i, \kappa' j} (\mathbf{q})-\delta_{\kappa \kappa'} \sum_{\kappa''}\hat{C}^{\mathrm{DD}}_{\kappa i, \kappa'' j} (\mathbf{q}=\mathbf{0}), 
\label{eq:long-range-dynamical-matrix} 
\end{eqnarray}
where $\hat{C}^{\mathrm{DD}}$ is the Fourier transformation of $C^{\mathrm{DD}}$, and the second term, which accounts for the ASR in the reciprocal space, should also be symmetric to keep the dynamical matrix $\tilde{C}^{\mathrm{DD}}$ Hermitian \footnote{An equivalent statement is that the sum in Eq.~(\ref{eq:long-range-dynamical-matrix}) can be performed on the first ($\hat{C}^{\mathrm{DD}}_{\kappa'' i, \kappa j} $) or second  ($\hat{C}^{\mathrm{DD}}_{\kappa i, \kappa'' j} $) atomic index.}:
\begin{align}
Q^{\mathrm{DD}}_{\kappa, ij} \equiv \sum_{\kappa''} \left[\hat{C}^{\mathrm{DD}}_{\kappa i, \kappa'' j} (\mathbf{q}=\mathbf{0}) -(i \leftrightarrow j) \right]=0.
\label{eq:symmetric-recip}
\end{align} 
The Born effective charges $Z^{*}$ can be factored out from $\hat{C}^{\mathrm{DD}}$:
\begin{equation} \label{eq:C-factor-out}
\hat{C}^{\mathrm{DD}}_{\kappa i, \kappa' j} (\mathbf{q}) =  \sum_{i^{\prime}j^{\prime}} Z^{*}_{\kappa, ii^{\prime}} Z^{*}_{\kappa',j j^{\prime}} \bar{C}^{\mathrm{DD}}_{\kappa i', \kappa' j'} (\mathbf{q}),
\end{equation}
and $\bar{C}^{\mathrm{DD}}(\mathbf{q})$, independent of $Z^{*}$, can be efficiently computed by Ewald summation of $ e^{i \mathbf{d}\cdot \mathbf{q}} V^{\mathrm{DD}}(0, \mathbf{R}) $ \cite{Gonze1997PRB10355}.

\subsection{Corrections for non-Hermiticity} \label{sec:dpcor}
However, closer examination of the long-range force constant ansatz in eq.~(\ref{eq:CDD-prev}) reveals a severe problem. $C_{\kappa i, \kappa' j}^{\mathrm{DD}}$ is symmetric with respect to the composite atomic {\em and } cartesian indices ($\kappa i \leftrightarrow \kappa' j$), a prerequisite for force constants, but not to cartesian indices alone, meaning that eq.~(\ref{eq:symmetric}) is not automatically satisfied term by term. In fact this leads to non-vanishing Eq.~(\ref{eq:symmetric}) and its reciprocal-space equivalent Eq.~(\ref{eq:symmetric-recip}), as well as a non-Hermitian dynamical matrix $\tilde{C}^{\mathrm{DD}}$ in general. This can also be understood with the observation that the tensors $Z^*_\kappa$ and $Q^{\mathrm{DD}}_\kappa$ are both constrained under the same site symmetry at $\kappa$. As $Z^*$ is in general not a symmetric tensor, the anti-symmetric $Q^{\mathrm{DD}}$ does not vanish either. Exceptions exist, e.g.\ when $C_{\kappa i, \kappa' j}^{\mathrm{DD}}$ becomes symmetric and Eq.~(\ref{eq:symmetric}) holds term by term, or when  $Q^{\mathrm{DD}}_{\kappa} $ vanishes due to point symmetry of site $\kappa$. Such special cases include, mutually non-exclusively:
\begin{itemize}
	\item binary semiconductors AB with two atoms per primitive cell and $Z^{*}_{\mathrm{A}} = - Z^{*}_{\mathrm{B}}$, including the rock salt, cesium chloride, and zinc blende crystal structures,
	\item solids with cubic symmetry on all sites and hence scalar Born effective charges $Z^{*}_{\kappa,ij} \propto \delta_{ij} $, e.g.\ the fluorite (AB$_2$) and perovskite (ABX$_3$ and AX$_3$) crystal structures,
	\item other structures with linearly related $Z^{*}$ tensors, e.g.\ wurtzite,
	\item site symmetry groups of all occupied Wyckoff positions have at least two perpendicular  rotation axes or mirror planes, including the $mmm$, $\bar{3}m$, $4/mmm$, $6/mmm$, $m\bar{3}$, $m\bar{3}m$ Laue classes. These groups guarantee symmetric second-order tensors (including $Z^{*}_{\kappa}$) and hence vanishing $Q^{\mathrm{DD}}_{\kappa}$.
\end{itemize}
Note that the ASR for $Z^{*}$, or the charge neutrality condition, $\sum_{\kappa} Z^{*}_{\kappa,ij} =0$, does not guarantee Eq.~(\ref{eq:symmetric}). Numerical tests, including using an independent DFPT code, also reveal ubiquitous non-Hermitian $\tilde{C}^{\mathrm{DD}}$ in low-symmetry polar semiconductors.

We would like to make a few general comments on FCMs. Symmetric FCMs may be derived naturally from a pairwise, translationally invariant potential $\sum_{a<b} E(r_{a,1}-r_{b,1},r_{a,2}-r_{b,2},r_{a,3}-r_{b,3})$. Asymmetry, if present in a FCM, has to come from many-body effects and will be canceled in the sum of Eq.~(\ref{eq:symmetric}) due to translational invariance.
We conclude that, except for special cases, Eq.~(\ref{eq:CDD-prev}) cannot be the second derivatives of a translationally invariant potential energy function and is not a valid force constant ansatz for {\em all} distances. This does not negate the DFPT procedure of Gonze and Lee, since the latter requires the total dynamical matrix $\tilde{D}$ to be Hermitian (or equivalent the total force constants to satisfy the ASR), not individual short- and long-range terms. However, the non-hermiticity issue is in conflict with our goal towards short-range force constants that satisfy the ASR.

As a remedy, we propose a new form of long-range FCM with real-space corrections $\Phi^{\mathrm{cor}}$,
\begin{align}
C^{\mathrm{DD, New}}_{\kappa i, \kappa' j}=C^{\mathrm{DD}}_{\kappa i, \kappa' j}+ \Phi^{\mathrm{cor}}_{\kappa i, \kappa' j}.
\end{align}
The short-ranged (see discussions later) FCMs $\Phi^{\mathrm{cor}}$ help $C^{\mathrm{DD, New}}$ satisfy both the long-range behavior of Eq.~(\ref{eq:CDD-prev}) and Hermiticity of Eq.~(\ref{eq:symmetric-recip}):
\begin{eqnarray}
& & \sum_{\kappa''} \left[\hat{C}^{\mathrm{DD,New}}_{\kappa i, \kappa'' j} (\mathbf{q}=\mathbf{0}) -(i \leftrightarrow j) \right] \nonumber \\   
&=&  \sum_{\kappa''} \left\{ \left[\hat{\Phi}^{\mathrm{cor}}_{\kappa i, \kappa'' j} (\mathbf{q}=\mathbf{0}) + \hat{C}^{\mathrm{DD}}_{\kappa i, \kappa'' j} (\mathbf{q}=\mathbf{0}) \right] -(i \leftrightarrow j) \right\} \nonumber \\
&=&  \sum_{\kappa'' } \left[ \sum_{\mathbf{R}} {\Phi}^{\mathrm{cor}}_{\kappa i, \kappa'' j} (0, \mathbf{R}) -(i \leftrightarrow j)   \right] +Q^{\mathrm{DD}}_{\kappa,ij} \nonumber \\
&=&\sum_{\kappa', \mathbf{R} \neq \kappa, 0} \left[ \Phi_{\kappa i, \kappa' j}^{\mathrm{cor}} (0, \mathbf{R}) - (i \leftrightarrow j)  \right]  +Q^{\mathrm{DD}}_{\kappa,ij} =0.
\label{eq:mod-linear-eq}
\end{eqnarray}
Here the Fourier transformation $\hat{\Phi}^\text{cor}$ does not include the on-site term $\Phi_{\kappa i, \kappa j}^{\mathrm{cor}} (0, 0)$. 
To properly account for crystal symmetry, we adopt a procedure similar to Section II of Part I. 
$\Phi^{\mathrm{cor}}$ is symmetrized using space group symmetry, {\em excluding} translational invariance, as
\begin{align}
\mathbf{\Phi}^{\mathrm{cor}} = \mathbb{C}^{\mathrm{spg}} \boldsymbol{\phi} ^{\mathrm{cor}}, \label{eq:mod-symmetry}
\end{align}
where $\mathbf{\Phi}^{\mathrm{cor}}$ is the one-dimensional list combining matrix elements of  representative $\Phi_{\kappa i, \kappa' j}^{\mathrm{cor}} (0, \mathbf{R})$ ($\kappa', \mathbf{R} \neq \kappa, 0$) excluding on-site terms, and $\boldsymbol{\phi}^{\mathrm{cor}}$ is the symmetry-reduced list of parameters. The matrix $\mathbb{C}^{\mathrm{spg}}$ is derived from symmetry of the interactions.
The on-site term is as usual dictated by the ASR:
\begin{align}
\Phi_{\kappa i, \kappa j}^{\mathrm{cor}} (0, 0)= -\sum_m \mathbb{B}^{\mathrm{ASR}}_{\kappa i j, m}  {\Phi}^{\mathrm{cor}}_m, \label{eq:mod-ASR}
\end{align}
where $\mathbb{B}^{\mathrm{ASR}} $ is a matrix that takes into account the crystal structure as discussed in Part I. $\Phi_{\kappa i, \kappa j}^{\mathrm{cor}} (0, 0)$ is not required to be symmetric. Instead,
to cancel the anti-symmetric $Q^{\mathrm{DD}}$ from Eqs.~(\ref{eq:mod-linear-eq},\ref{eq:mod-symmetry}), we solve for
\begin{align}
\sum_m \left[\mathbb{B}^{\mathrm{ASR}}_{\kappa i j, m} - (i \leftrightarrow j) \right] (\mathbb{C}^{\mathrm{spg}} \boldsymbol{\phi} ^{\mathrm{cor}})_m = -Q^{\mathrm{DD}}_{\kappa,ij}.
\label{eq:mod-linear-eq-symmetry}
\end{align}
This is an under-determined set of linear equations for unknowns $ \boldsymbol{\phi} ^{\mathrm{cor}}$, as long as the interaction distance cutoff for the correction FCMs is large enough to include more unknown parameters than known asymmetric elements of $Q^{\mathrm{DD}}$ (at most 3 times number of symmetrically distinct atoms). To keep the corrections simple, one may include as few short-range pairs as possible in $\Phi^{\mathrm{cor}}$. Different choices in $\Phi^{\mathrm{cor}}$ will be compensated by $\Phi^{\mathrm{SR}}$, so the total corrected dynamical matrix $ \tilde{C}^{\text{SR}} + \tilde{C} ^{\mathrm{DD}} + \tilde{C} ^{\mathrm{cor}}$ is not affected. Here $\tilde{C} ^{\mathrm{cor}}$ is the Fourier transformation of $\Phi^{\mathrm{cor}}$ with a simple real-space sum, just like $\tilde{C}^{\text{SR}}$. In this work we considered correctional interactions within the first coordination shell of any symmetrically distinct atom $\kappa$ to cancel $Q^{\mathrm{DD}}_\kappa$. The small under-determined problem was solved exactly with compressive sensing (no approximation necessary). Finally, the full set of correctional force constants can be obtained from Eqs.~(\ref{eq:mod-symmetry}, \ref{eq:mod-ASR}). Overall our correction scheme has a simple physical picture: keeping the dipole-dipole force constant ansatz beyond the first coordination shell, while satisfying the acoustic sum rule by modification to the interactions within the first shell.

\begin{figure*}[htp]
	\includegraphics[width = 0.85 \linewidth]{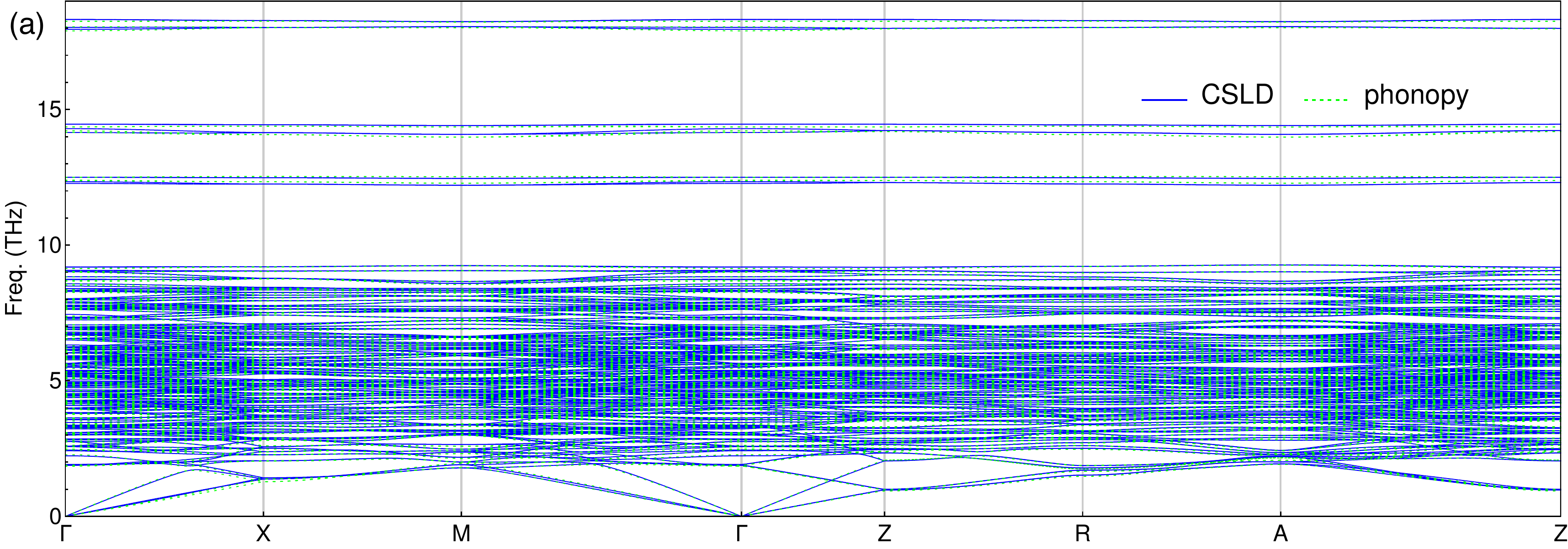}
	\includegraphics[width = 0.42 \linewidth]{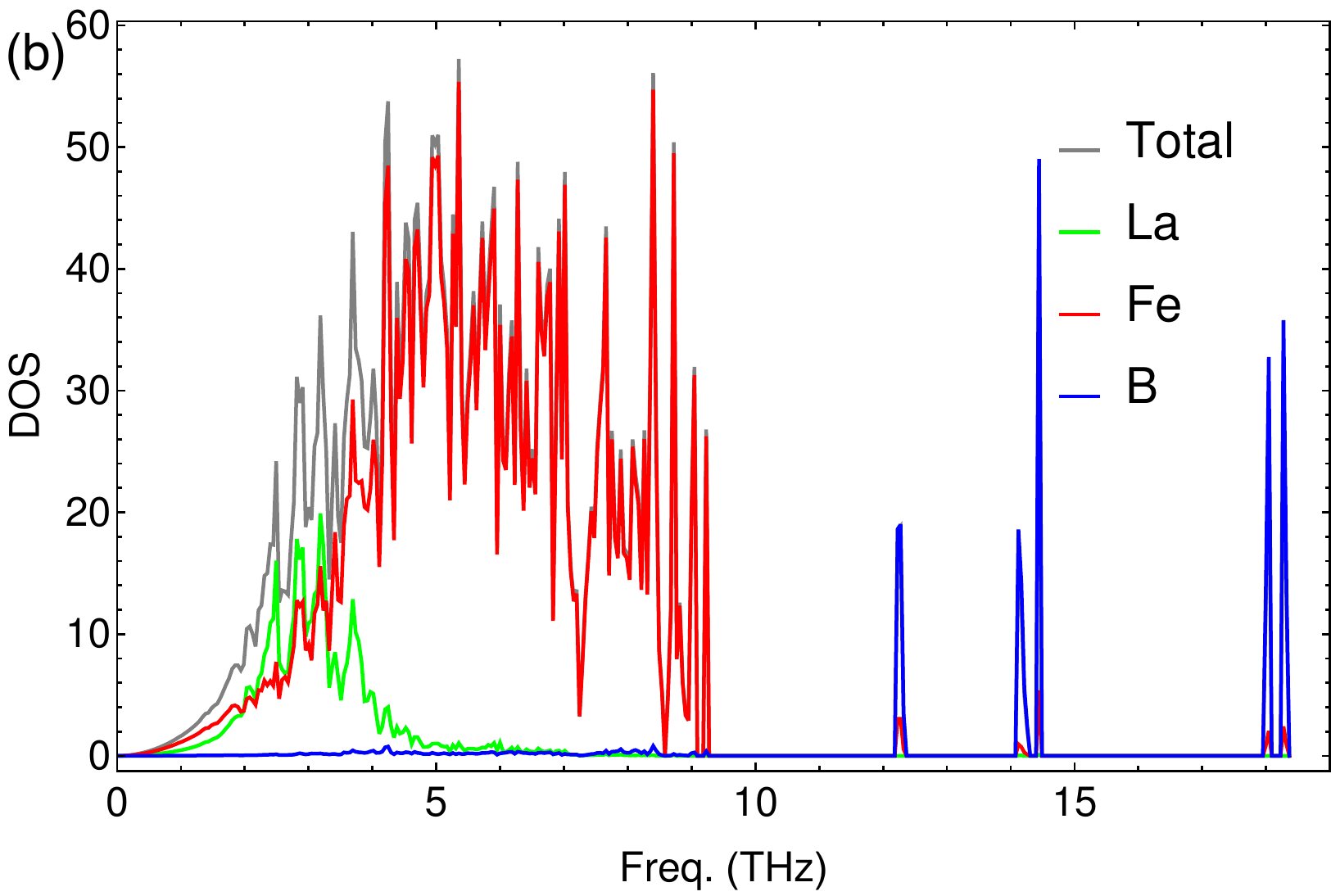}
	\includegraphics[width = 0.42 \linewidth]{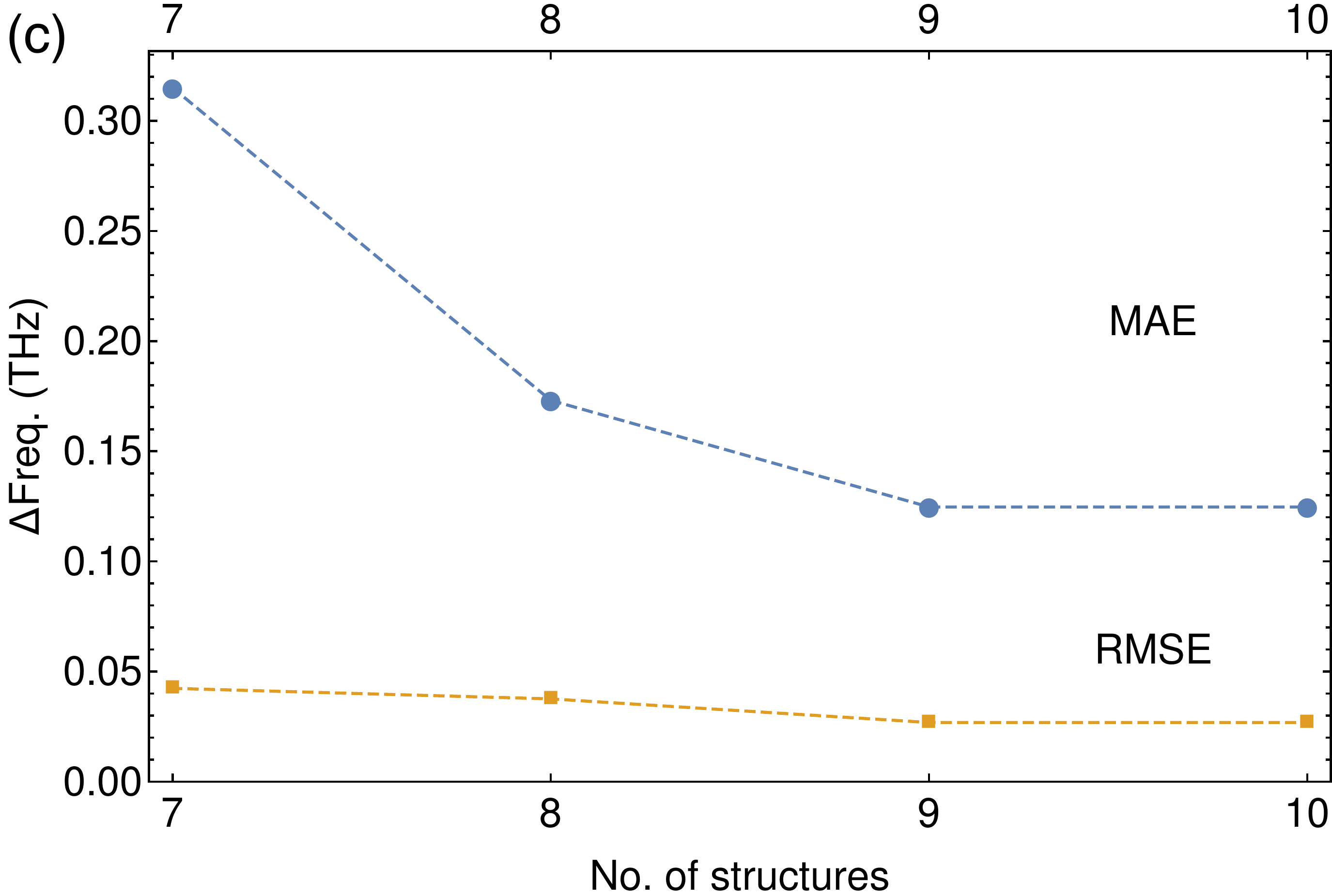}
	\caption{Phonon calculations of \ce{La2Fe14B}: (a) dispersion curves using CSLD with 10 supercell calculations (solid lines), and phonopy with 30 supercell calculations (dashed lines); (b) total and partial DOS using CSLD; and (c) root-mean-square (RMSE) and maximum absolute (MAE) errors of the CSLD band energies compared to phonopy vs.\ the number of CSLD training structures. The special points are $\Gamma$ (0,0,0); A (.5, .5, .5); M (.5, .5, 0); R (0, .5, .5); X (0, .5, 0); Z (0, 0, .5).}
	\label{fig:LaFeB}
\end{figure*}

\section{Results}
All calculations followed the same computational settings as Part I. DFT calculations adopted the  Perdew-Becke-Ernzerhof (PBE) version of the generalized gradient approximation (GGA) \cite{Perdew1996PRL3865} except for \ce{CeO2}, for which GGA+$U$ \cite{Anisimov1991PRB943} with  $U=5$ eV and the HSE06 hybrid method \cite{Krukau2006JCP224106} were used. While fittings of FCM can be performed using total energies if accurate Hellmann-Feynman forces are not available\cite{Soderlind2015SR15958}, in this work we fit forces. Fittings of pair force constant matrices were performed together with third order force constant tensors of the form $\Phi_{a,a,a}$ (one-atom anharmonicity) and $\Phi_{a,a,b}$ (nearest-neighbor pair anharmonicity) in order to increase the fitting accuracy, similar to the use of plus/minus displacement in the direct method. These third-order terms typically decrease the relative fitting error from 1--3\% to less than 1\%.  All atoms in the supercell calculations were independently displaced in a random direction  by 0.01 \AA\ away from  equilibrium, in contrast to the direct method, which usually moves one atom in one direction at a time. In polar semiconductors, the $Z^{*}_{\kappa,ij}$ and  $\epsilon_{ij}$ tensors were computed using DFPT and the PBE functional.

\subsection{Metallic solid}
The first case study is the rare-earth alloy \ce{Nd2Fe14B} (NdFeB), one of the most widely used permanent magnet.\cite{Croat1984APL148,Herbst1991RMP819} However, our GGA calculation failed to converge with satisfactory numerical precision due to the $4f$-electrons of neodymium. Instead, we studied \ce{La2Fe14B} as a model system for NdFeB by replacing Nd with La. 
The crystal structure is tetragonal with space group $P4_2/mnm$ and has 68 atoms in the unit cell. Spin-polarized GGA calculations were performed in a $2\times 2 \times 1$ supercell with 272 atoms. For comparison, the phonopy program \cite{Togo2015SM1} was also used, which adopts the conventional direct method for force constant computation. 

Fig.~\ref{fig:LaFeB} shows the obtained phonon dispersion and density of states (DOS) of \ce{La2Fe14B}. The phonon spectra feature acoustic modes with not only iron but also significant rare earth contributions due to the large mass of the latter. The optical phonons are clearly divided into iron-dominated low-frequency modes and high-frequency ones, which are well separated from other modes and mainly attributed to the light boron atoms. As shown in Fig.~\ref{fig:LaFeB}a, CSLD results with 10 supercell calculations are in excellent agreement with those using phonopy and 30 calculations. The difference in CSLD and phonopy band energies are shown in Fig.~\ref{fig:LaFeB}c versus the number of supercell structures used in CSLD fitting. Reasonable convergence is reached with as few as 8 structures. 

\subsection{Polar semiconductors and insulators}

\subsubsection{Separating short-range force} \label{sec:NaCl}
\begin{figure}[htp]
	\includegraphics[width = 0.67 \linewidth]{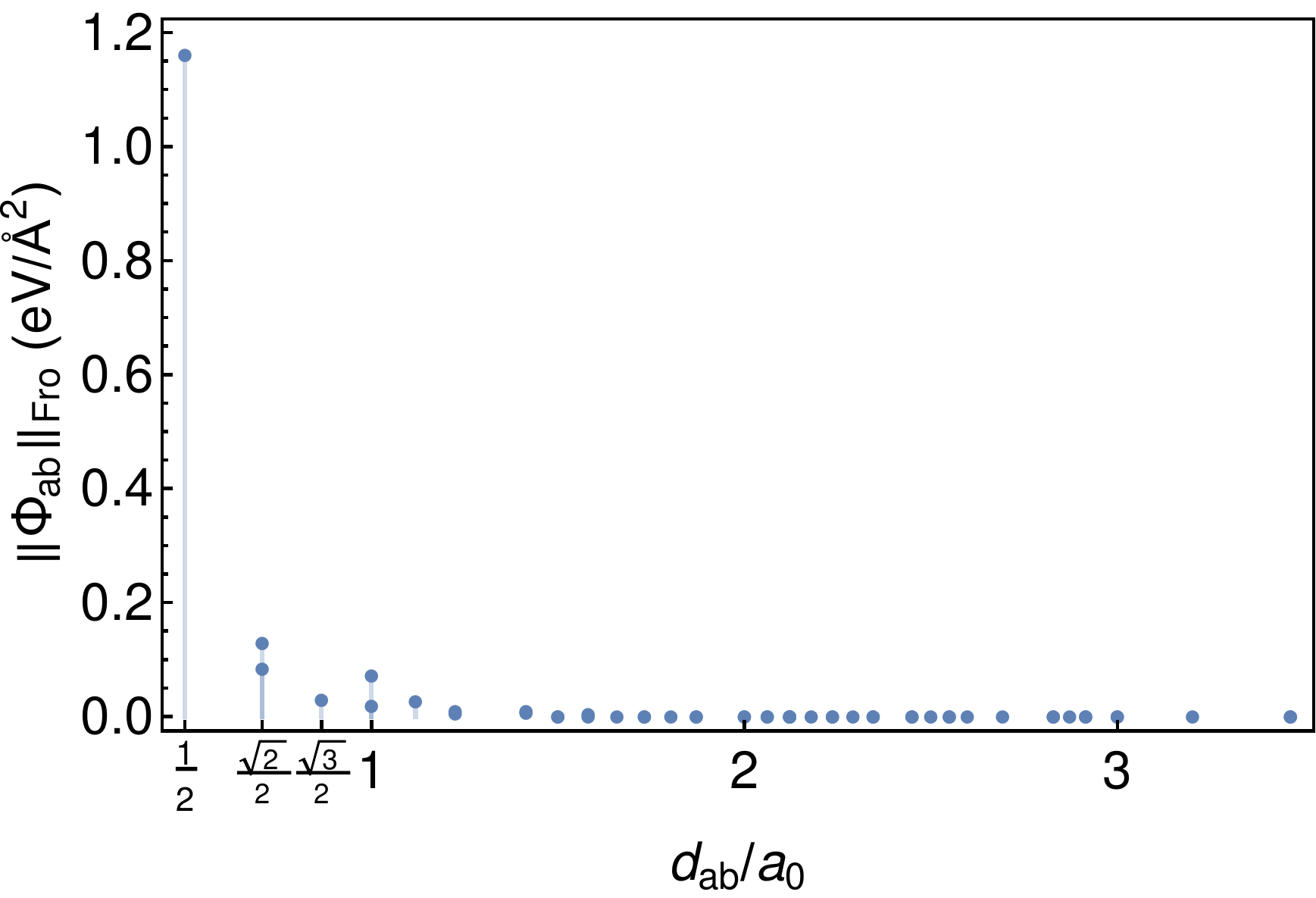}
	\includegraphics[width = 0.67 \linewidth]{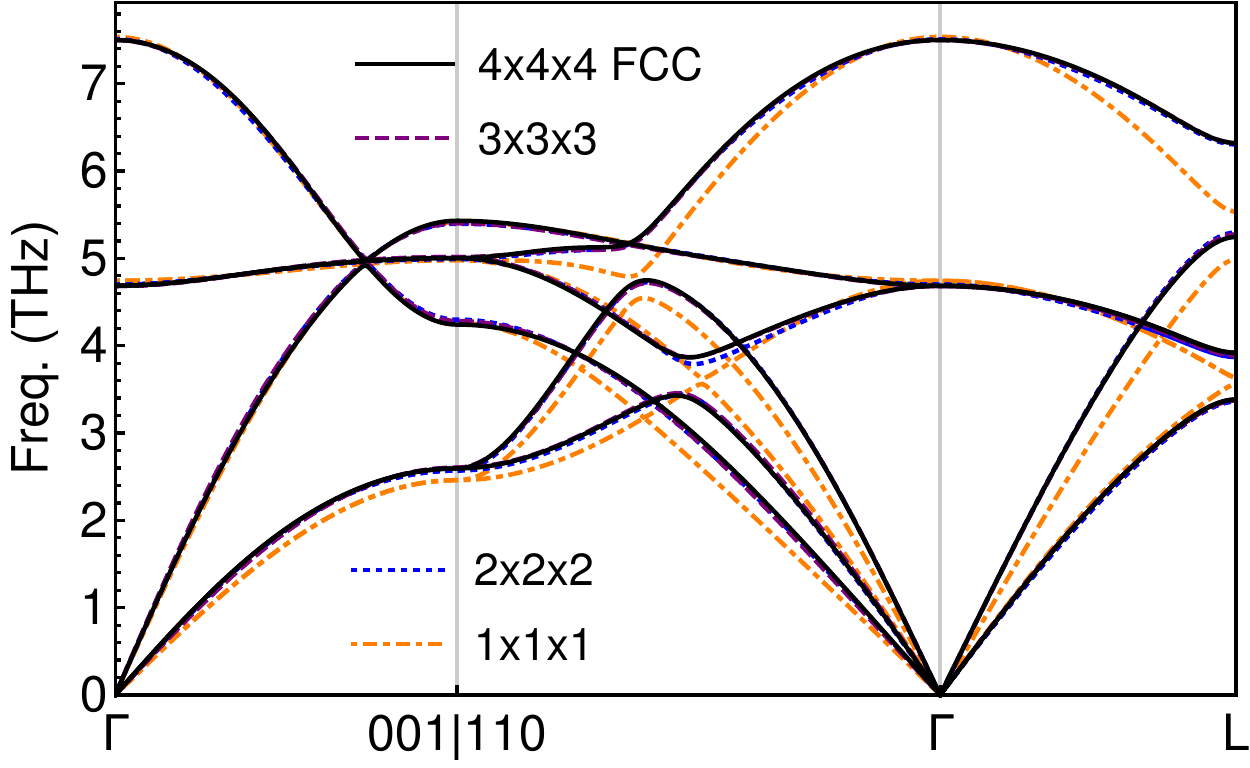}
	\includegraphics[width = 0.67 \linewidth]{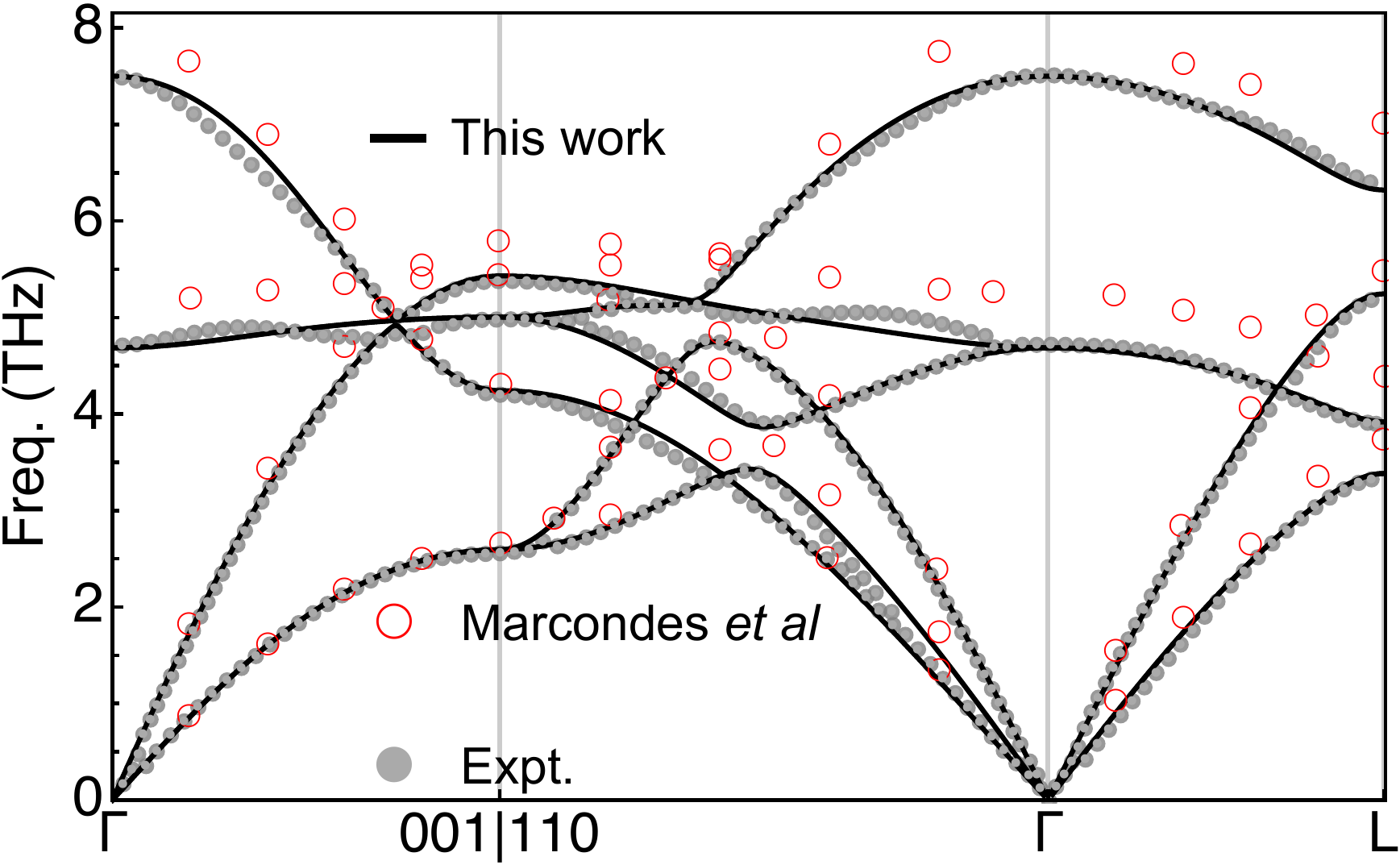}
	\caption{CSLD results on NaCl in $n \times n \times n$ fcc cells. Upper: Frobenius norm of the short-range force matrix  $\mathbf{\Phi}^{\text{SR}}$  
		vs.\ distance 
		($n=4$); middle: phonon dispersion ($n=1\textup{--}4$); bottom: comparison of converged phonon spectrum ($n=4$) with inelastic neutron scattering data from Ref.~\onlinecite{Raunio1969PR1496} (red circles) and previous PBE calculations by Marcondes {\it et al} from Ref.~\onlinecite{Marcondes2018SSC11}.}
	\label{fig:NaCl}
\end{figure}
In the test case of NaCl (as well as the next case of ceria), as discussed previously, both ions possess cubic point group symmetry and the dipole-dipole non-Hermitian correction is not necessary. Fig.~\ref{fig:NaCl} shows the results  with one single supercell calculation. As shown in Fig.~\ref{fig:NaCl}a, the dominant short-range interaction is the nearest-neighbor Na-Cl pair separated by $0.5 a_0$ (half lattice constant). Other short-range force constants are one order of magnitude smaller and practically vanish beyond $1.5 a_0$ according to the CSLD fitting, confirming that the residual interactions other than long-range electrostatics are indeed short-ranged, in agreement with similar observation achieved through reciprocal space range-separation \cite{Gonze1994PRB13035}. As a result, the calculated phonon dispersion (Fig.~\ref{fig:NaCl}b) is reasonably accurate using the small $1\times 1 \times 1$ conventional face-centered cubic (fcc) cell, which allows fitting of the shortest pairs, and well converged with the $2\times 2 \times 2$ fcc cell. Compared to our previous result (supplemental material of Ref.~\onlinecite{Zhou2014PRL185501}) using the Parlinski interpolation scheme, the current dispersion is smooth and free of the roughness near $\Gamma$. As shown in the bottom part of Fig.~\ref{fig:NaCl}, the acoustic modes are in good agreement with experiment\cite{Raunio1969PR1496},  while the optical modes are underestimated, due to PBE's insufficient description of van der Waals interactions \cite{Marcondes2018SSC11}. This serves as a cautionary tale that phonon calculations are sensitive to the exchange-correlation functional used, and appreciable errors may arise in ``trivial'' text-book systems like rock salt. With the same PBE functional, our results are nearly identical to previous ones \cite{Marcondes2018SSC11}.

\subsubsection{Combining hybrid functional calculations}

Hybrid functionals, the rung above (semi)-local approximations in Jacob's ladder for DFT, typically offer systematic improvement in electronic structure and mechanical properties. Hybrid-DFT is known for much better agreement with experimental phonon dispersion over GGA in e.g.\ \ce{CeO2}\cite{Wang2013PRB024304}. However, hybrid-DFT calculations as implemented in most DFT codes for periodic systems are considerably more expensive than standard DFT, particularly for large unit cells. Here we show that high-quality results may be obtained by combining conventional DFT calculations in large supercells and hybrid functional calculations in relatively small supercells. The short range force constants in Eq.~(\ref{eq:SR}) are decomposed into 
\begin{align} \label{eq:SR-hybrid}
\Phi_{ab}^{\text{SR}} = \tilde\Phi_{ab}^{\text{SR}} + \Delta \Phi_{ab}^{\text{SR}},
\end{align}
where $\Phi^{\text{SR}}$ and $\tilde\Phi^{\text{SR}}$  are the force constant matrices according to the more accurate but expensive (e.g.\ hybrid-DFT) and the less expensive (e.g.\ GGA) computational approaches, respectively, while $\Delta \Phi^{\text{SR}}$ is the difference. Assuming that the $\tilde\Phi^{\text{SR}}$ values are a reasonable approximation, one might expect  $\Delta \Phi^{\text{SR}}$ to be even more sparse. The strategy
is to first fit the normal GGA short-range force constants $\tilde\Phi_{ab}^{\text{SR}}$, and then obtain the hybrid corrections to GGA according to Eq.~(\ref{eq:SR-hybrid}).

\begin{figure}[hpt]
	\includegraphics[width = 0.89 \linewidth]{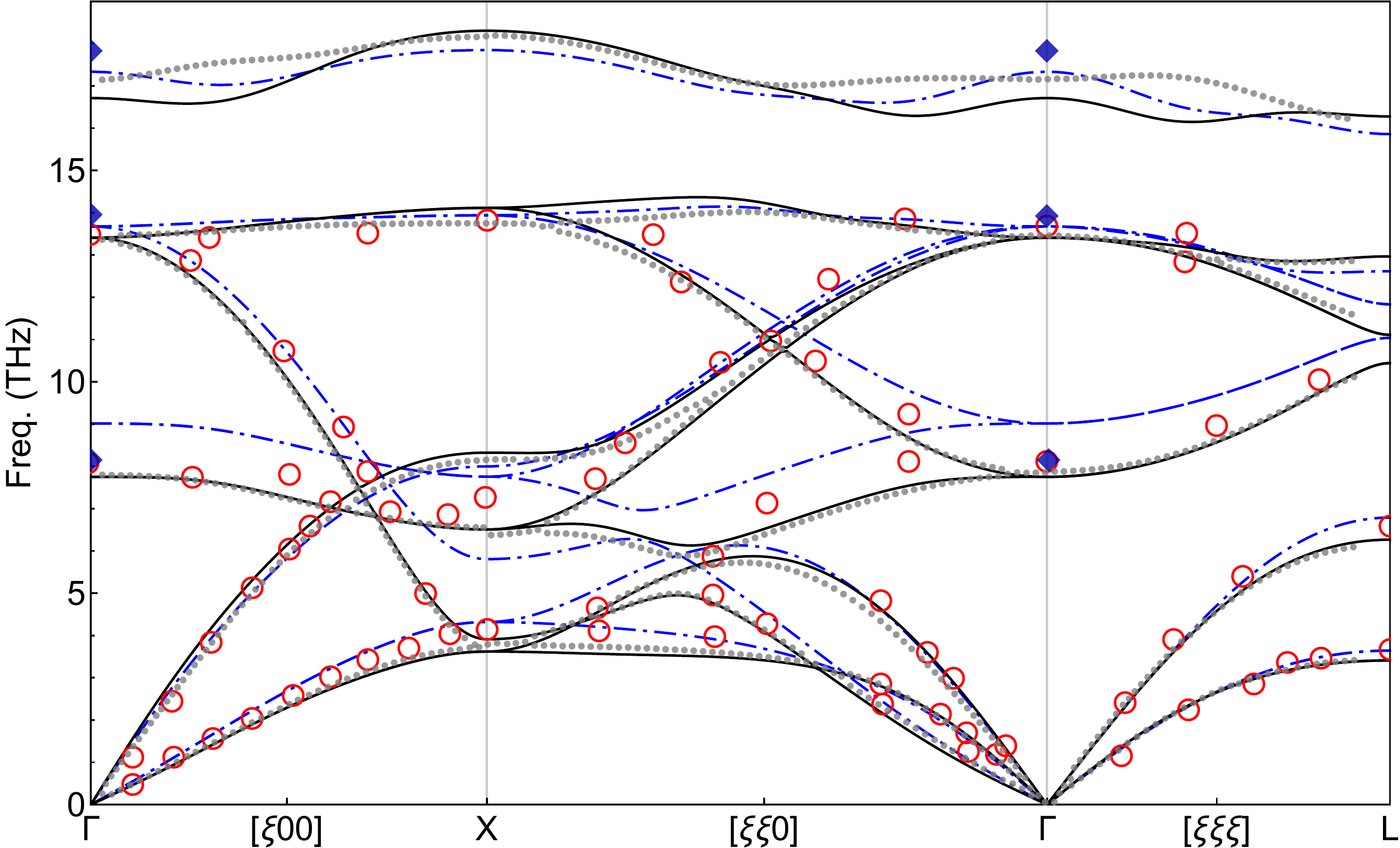}
	\caption{CSLD phonon dispersion for \ce{CeO2} from GGA+$U$ (dot-dashed lines) and HSE06 hybrid (solid) calculations. For comparison, data points from the literature were shown: red open circles: neutron scattering data  (Ref.~\onlinecite{Clausen1987CeO2}); solid diamonds at the $\Gamma$ point: Raman and infrared data (Ref.~\onlinecite{Mochizuki1982CeO2}); gray dotted lines: full phonon calculations by Wang {\it et al} (Ref.~\onlinecite{Wang2013PRB024304})}
	\label{fig:CeO2-neutral-band}
\end{figure}

One GGA+$U$ calculation for \ce{CeO2} was carried out in the $3\times 3 \times 3$ fcc cell following the same procedure as in the previous NaCl example.  Similar to the NaCl results, the short-range force constants are indeed found to be short ranged: they are practically identical compared to those using the $4\times 4 \times 4$ cell and very close to the $2\times 2 \times 2$ fcc cell (not shown). Relatively minor difference in phonon dispersion was found using GGA+$U$ or GGA (not shown). To get the hybrid corrections $\Delta \Phi_{ab}^{\text{SR}}$, two HSE06 hybrid calculations were performed in a $2\times 2 \times 2$ supercell of the primitive cell of ceria (8 formula units) and fitted with CSLD.  The hybrid-corrected dispersion (with HSE06 modifications to GGA+$U$ only on the nearest- and 2nd-nearest-neighbor force constants) is shown in Fig.~\ref{fig:CeO2-neutral-band}. The phonon dispersion is almost identical with the results of Ref.~\onlinecite{Wang2013PRB024304} using HSE06 calculations, phonopy and the direct method, shown in Fig.~\ref{fig:CeO2-neutral-band} as gray dotted lines, except for the LO modes near the $\Gamma$ point. The difference in long-wavelength LO modes may come from the different treatment of LO-TO splitting (interpolation of the non-analytic correction versus our separated treatment of short-range and long-range interactions in Section \ref{sec:separation}) and/or different $Z^*$ and $\epsilon^\infty$ parameters used. Similar to Ref.~\onlinecite{Wang2013PRB024304}, our results are in good agreement with experimental data, shown in Fig.~\ref{fig:CeO2-neutral-band} as large red circles and diamonds. This result suggest that the predominant effects of hybrid functional on lattice dynamics on very short-ranged in CeO$_2$.  Note that 
we kept the $Z^*$ and $\epsilon^\infty$ values were from GGA+$U$ rather than HSE06 calculations.

\begin{figure*}[hpbt]
	\includegraphics[width = 0.42 \linewidth]{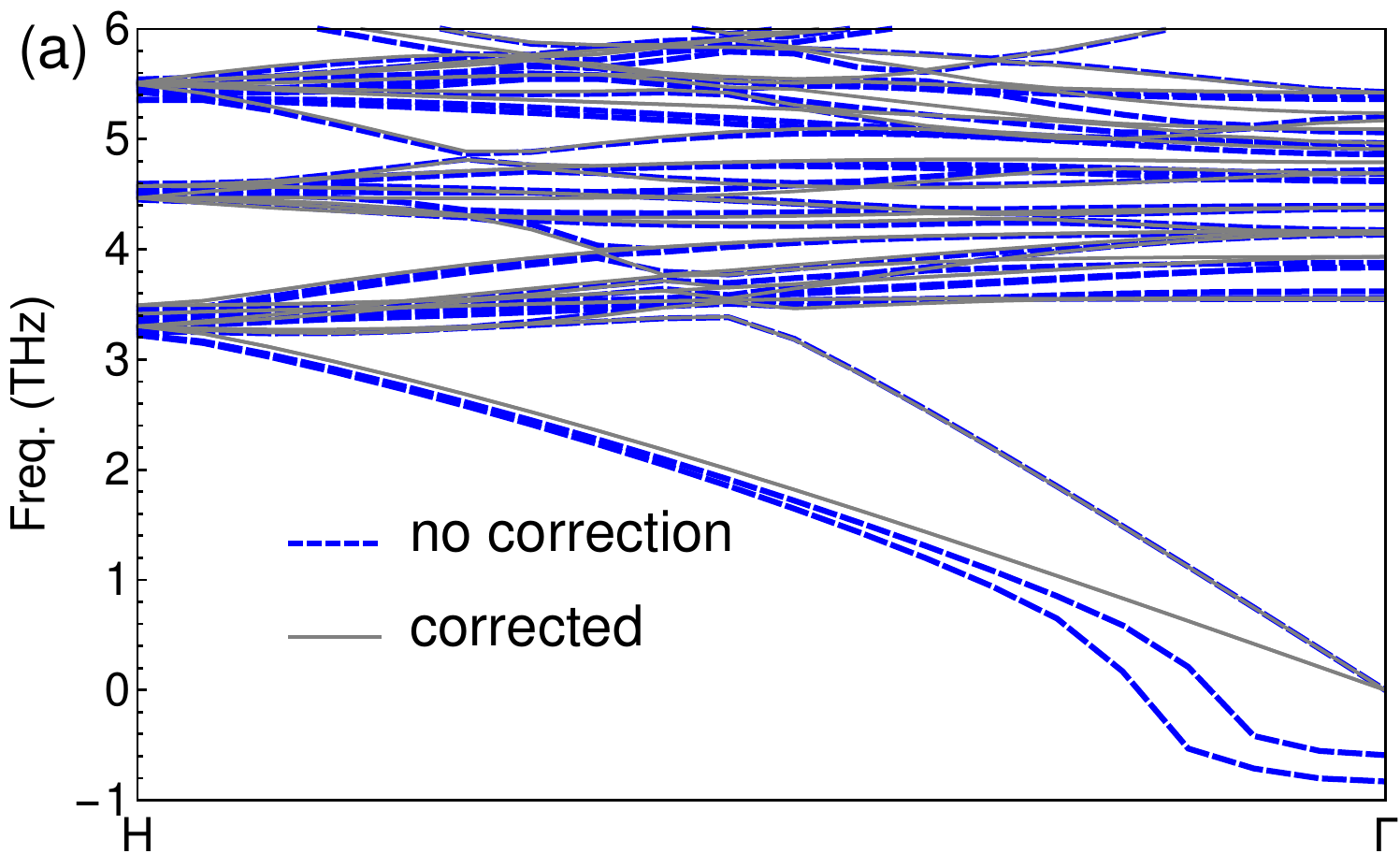}
	\includegraphics[width = 0.42 \linewidth]{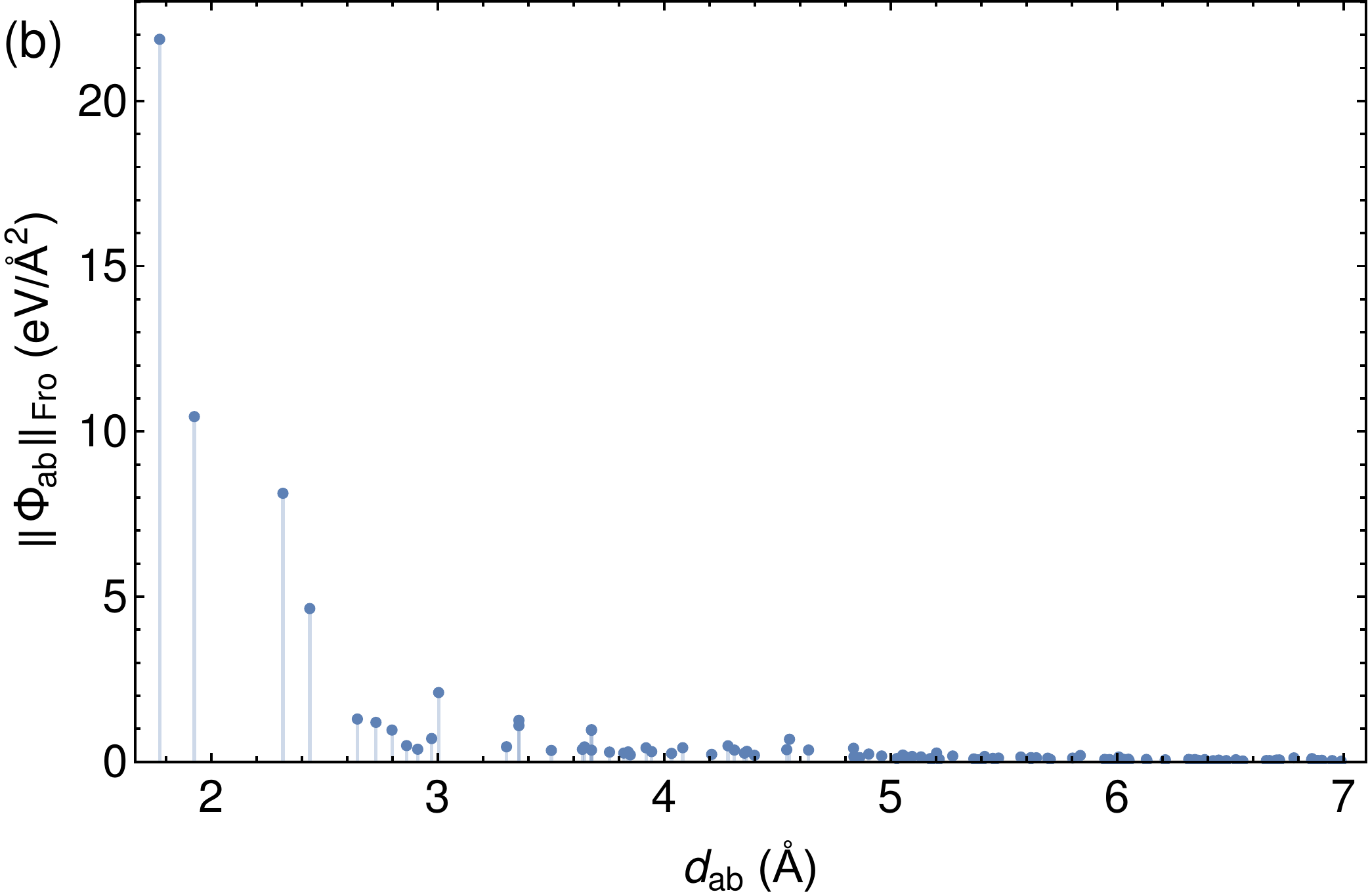}	
	\includegraphics[width = 0.42 \linewidth]{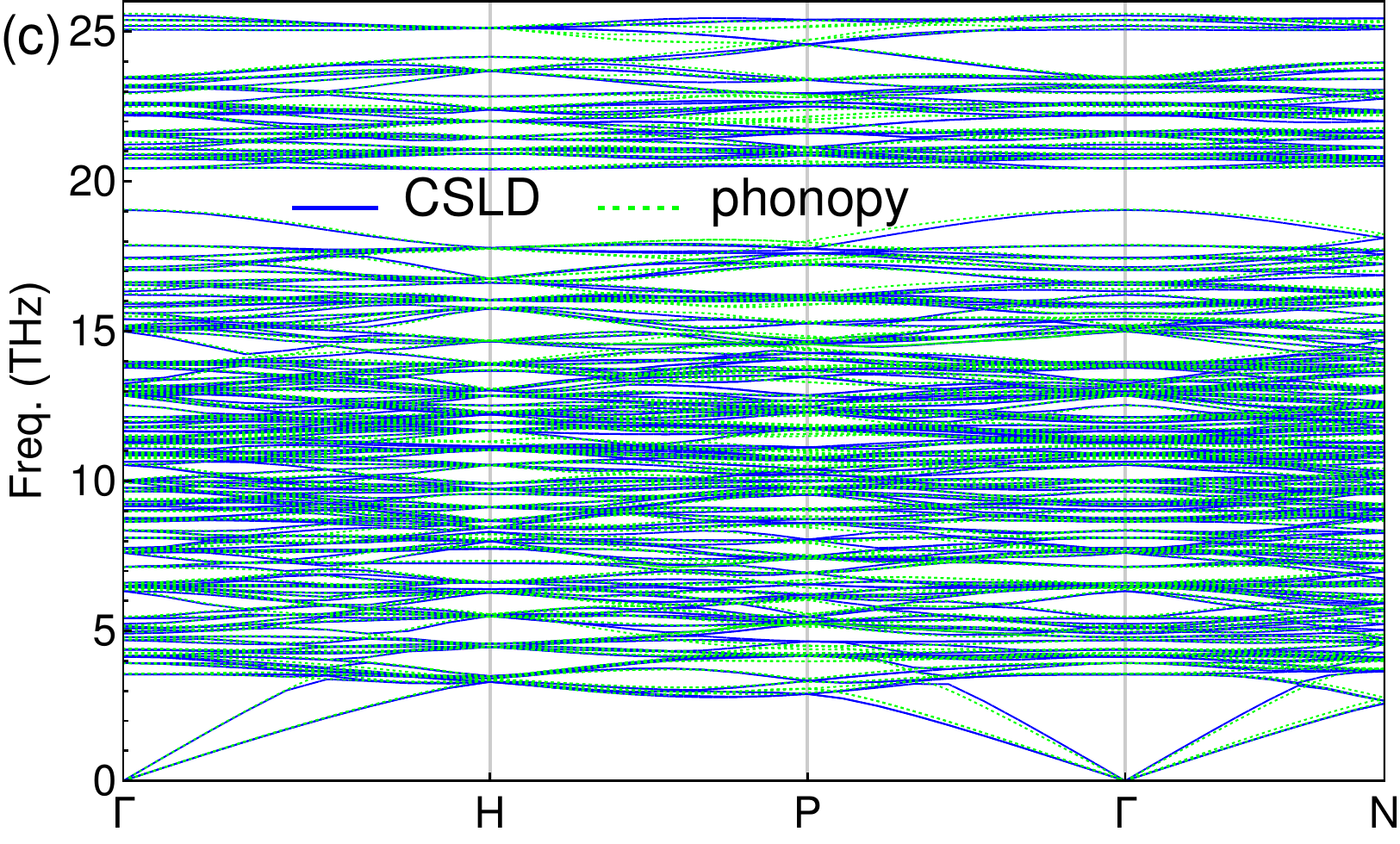}
	\includegraphics[width = 0.42 \linewidth]{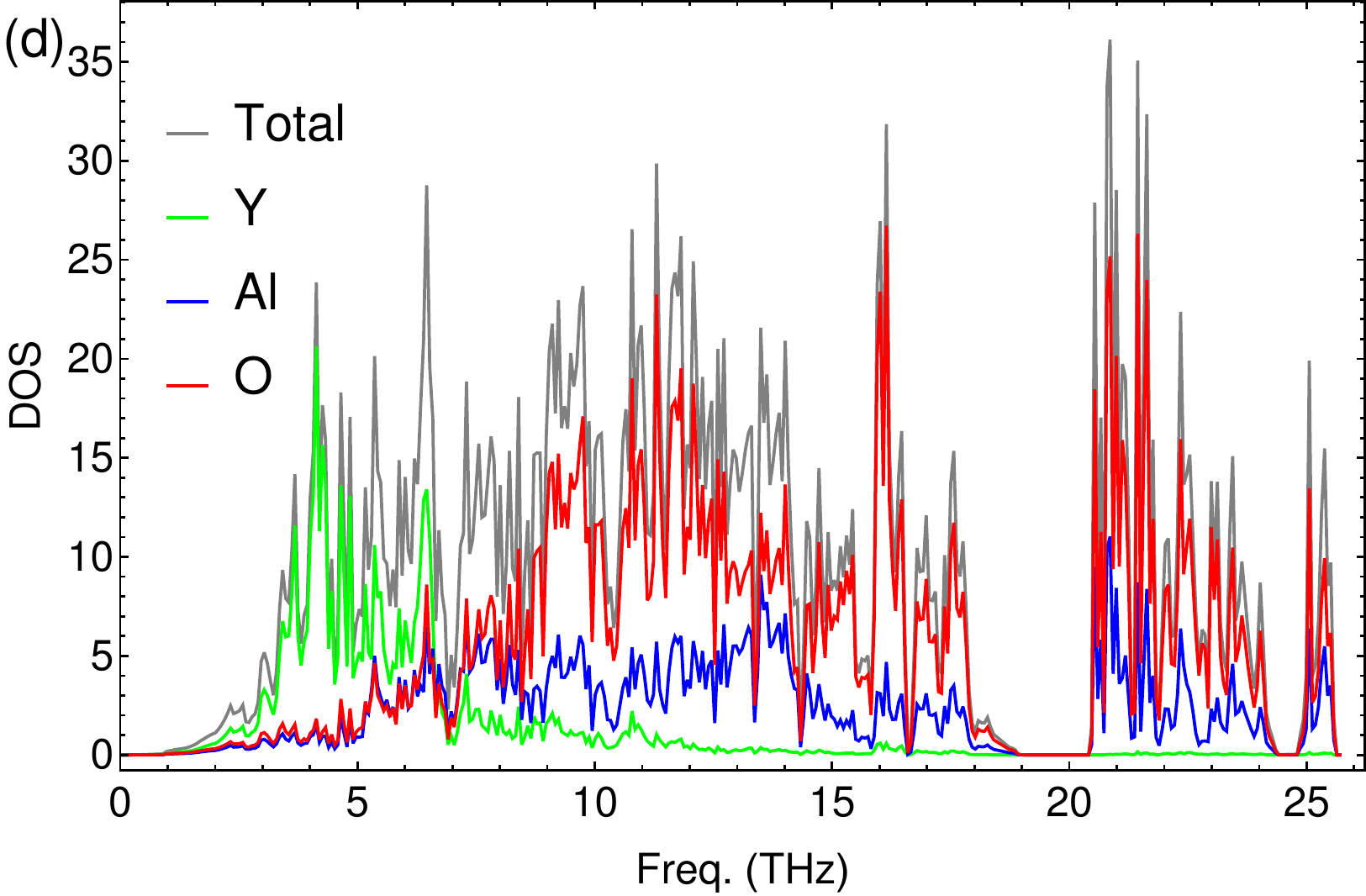}
	\caption{CSLD phonon calculations for \ce{Y3Al5O12}: (a) phonon dispersion before and after the correction, and with the correction, (b) Frobenius norm of the short-range force matrix  $\mathbf{\Phi}^{\text{SR}}$ vs.\ interaction distance, (c) dispersion curves compared with phonopy results, and (d) total and partial phonon DOS.}
	\label{fig:YAG}
\end{figure*}
\subsubsection{Dipole-dipole non-Hermitian correction in YAG}
Now we demonstrate the dipole-dipole non-Hermitian correction on yttrium aluminum garnet (YAG, \ce{Y3Al5O12}), a solid-state laser material with space group $I a \bar3 d$ and 80 atoms in the body-centered cubic (bcc) primitive cell. Despite the cubic space group, the point groups of the ions are lower and give rise to non-scalar Born effective charges and a non-Hermitian dipole-dipole dynamical matrix, as discussed in Section \ref{sec:dpcor}. In particular, the anti-symmetric tensor $Q^{\mathrm{DD}}$ was found to vanish only on the yttrium sublattice with Wyckoff position $24c$ (site symmetry $2.2\ 2$), not on Al(1) on $24d$ (symmetry $-4..$), Al(2) on $16a$ ($.-3.$) or oxygen on $96h$ (1). The ill-defined dynamical matrix becomes particularly erratic on the acoustic branches near the zone center, as shown in  Fig.~\ref{fig:YAG}a with the real part of the eigenvalues (dashed lines). 

With the corrected dipole-dipole FCM introduced in Section \ref{sec:dpcor}, a Hermitian dynamical matrix is recovered together with the vanishing acoustic modes near $\Gamma$ (solid lines in Fig.~\ref{fig:YAG}a). Compared to NaCl, the residual short-range force constants are one order of magnitude larger (Fig.~\ref{fig:YAG}b), suggesting significantly more covalent nature of the bonding in the \ce{Y3Al5O12} oxide.
Fig.~\ref{fig:YAG}c shows the CSLD calculated phonon dispersion (solid lines)  using 3 training structures in the conventional bcc cell (160 atoms), again in good agreement with the direct method using 11 training structures as implemented in phonopy. The total and projected DOS plots are displayed in Fig.~\ref{fig:YAG}d.

\section{Discussions and Conclusions}
Now we address the important question of convergence of phonon calculations using CSLD. First, how large does the supercell have to be? The examples presented suggest that the (short-range) force constants practically vanish at a distance of 7--9 \AA. More generally, one may monitor the magnitude of the force constants versus distance (e.g.\ Fig.~\ref{fig:NaCl}). If the obtained force constants of the longest distance are appreciable, then the used supercell is too small. If computational costs are a concern, mix-and-matching supercells of different shape and/or aspect-ratio is completely compatible with CSLD, as possibly problematic long-range electrostatic forces are already taken out. Second, how many training supercell structures are needed? A main advantage of CSLD is that it requires fewer structures than the direct method.  If the crystal symmetry is very low or the primitive cell is huge, the reduction can be more significant, as the number of required supercell structures is capped at a finite $\sim 6 N_a $. In general, one should keep a certain number of data points as a prediction set apart from the training set \footnote{see Section III of Part I. The use of a prediction set is common practice in supervised machine learning methods.}, monitor the prediction error, and add training structures if required. The examples show empirically 1/3--1/4 of the requirement number of supercell calculations in the direct method are enough for force prediction error of $\lesssim 1$\% for semiconductors and $\sim 2$\% for metals. Finally and broadly applicable to any first-principles phonon calculations, care should be taken to ensure convergence with respect to computational settings such as the number of $k$-points and energy cut-off \cite{Grabowski2007PRB024309}.

In conclusion, the compressive sensing lattice dynamics method is applied to calculate the phonon spectrum of a few metallic and semiconductor solids. Through a few case studies, we show that the CSLD method is particularly efficient for systems with large supercells and requires fewer training supercell structures than the direct method by effectively taking advantage of the ``short-sightedness'' of the atomic interactions. In polar semiconductors, this is achieved after the long-range Coulomb forces and force constants are separated, and the non-Hermitian dynamical matrix of the dipole-dipole force constant matrix is corrected with a few short-range terms.

\begin{acknowledgments}
The work of F.Z., B.S.\ and D.\AA.\ was supported by the Critical Materials Institute, an Energy Innovation Hub funded by the U.S. Department of Energy, Office of Energy Efficiency and Renewable Energy, Advanced Manufacturing Office, and performed under the auspices of the U.S. Department of Energy by Lawrence Livermore National Laboratory under Contract DE-AC52-07NA27344. V.O. acknowledges support from the U.S. Department of Energy, Office of Science, Basic Energy Sciences, under Grant DE-FG02-07ER46433, and computational resources from the National Energy Research Scientific Computing Center, which is supported by the Office of Science of the U.S. Department of Energy under Contract No. DE-AC02-05CH11231.
\end{acknowledgments}

\newpage
\clearpage
\onecolumngrid
\appendix
\renewcommand\thefigure{S\arabic{figure}}    
\setcounter{figure}{0}

\end{document}